\documentstyle[12pt,epsfig]{article}

\def\baselinestretch{1.2}
\def\href#1#2{#2}  

\newcommand{\norm}[1]{\raise.3ex\hbox{:} #1 \raise.3ex\hbox{:}\,}

\newfont{\Bbb}{msbm10 scaled 1200}     
\newcommand{\mathbb}[1]{\mbox{\Bbb #1}}

\def\cN{{\cal N}}

\newcommand{\beq}{\begin{equation}}
\newcommand{\eeq}{\end{equation}}
\newcommand{\beqar}{\begin{eqnarray}}
\newcommand{\eeqar}{\end{eqnarray}}
\def\appendix{{\newpage\section*{Appendix}}\let\appendix\section%
        {\setcounter{section}{0}
        \gdef\thesection{\Alph{section}}}\section}

\newcommand{\be}{\begin{equation}}
\newcommand{\ee}{\end{equation}}
\newcommand{\eel}[1]{\label{#1}\end{equation}}
\newcommand{\bea}{\begin{eqnarray}}
\newcommand{\eea}{\end{eqnarray}}
\newcommand{\eeal}[1]{\label{#1}\end{eqnarray}}
\newcommand{\baq}{\begin{equation}\begin{array}{rcl}}
\newcommand{\eaq}{\end{aryray}\end{equation}}
\newcommand{\eaql}[1]{\end{array}\label{#1}\end{equation}}
\newcommand{\beac}{\begin{equation}\begin{array}{rcl}}
\newcommand{\eeacn}[1]{\end{array}\label{#1}\end{equation}}
\newcommand{\ba}{\begin{array}}
\newcommand{\ea}{\end{array}}

\newcommand{\journal}[4]{{\rm #1~}{\bf #2}\,(#3)\,#4}

\newcommand{\ijmp}{\journal {Int. J. Mod. Phys.}}

\newcommand{\pr}{\journal {Phys. Rev.}}

\newcommand{\prl}{\journal {Phys. Rev. Lett.}}

\newcommand{\rmp}{\journal {Rev. Mod. Phys.}}

\newcommand{\cmp}{\journal {Comm. Math. Phys.}}
\newcommand{\cqg}{\journal {Class. Quantum Grav.}}

\newcommand{\np}{\journal {Nucl. Phys.}}
\newcommand{\pl}{\journal {Phys. Lett.}}
\newcommand{\mpl}{\journal {Mod. Phys. Lett.}}

\newcommand{\ptp}{\journal {Progr. Theor. Phys.}}
\newcommand{\nc}{\journal {Nuovo Cim.}}

\newcommand{\grg}{\journal {Gen. Rel. Grav.}}

\catcode`\@=11

\@addtoreset{equation}{section}
\catcode`\@=12

\def\noj#1,#2,{{\bf #1} (19#2)\ }
\def\jou#1,#2,#3,{{\sl #1\/ }{\bf #2} (19#3)\ }
\def\ann#1,#2,{{\sl Ann.\ Physics\/ }{\bf #1} (19#2)\ }
\def\cmp#1,#2,{{\sl Comm.\ Math.\ Phys.\/ }{\bf #1} (19#2)\ }
\def\ma#1,#2,{{\sl Math.\ Ann.\/ }{\bf #1} (19#2)\ }
\def\jd#1,#2,{{\sl J.\ Diff.\ Geom.\/ }{\bf #1} (19#2)\ }
\def\invm#1,#2,{{\sl Invent.\ Math.\/ }{\bf #1} (19#2)\ }
\def\cq#1,#2,{{\sl Class.\ Quantum Grav.\/ }{\bf #1} (19#2)\ }
\def\cqg#1,#2,{{\sl Class.\ Quantum Grav.\/ }{\bf #1} (19#2)\ }
\def\ijmp#1,#2,{{\sl Int.\ J.\ Mod.\ Phys.\/ }{\bf A#1} (19#2)\ }
\def\jmphy#1,#2,{{\sl J.\ Geom.\ Phys.\/ }{\bf #1} (19#2)\ }
\def\jams#1,#2,{{\sl J.\ Amer.\ Math.\ Soc.\/ }{\bf #1} (19#2)\ }
\def\grg#1,#2,{{\sl Gen.\ Rel.\ Grav.\/ }{\bf #1} (19#2)\ }
\def\mpl#1,#2,{{\sl Mod.\ Phys.\ Lett.\/ }{\bf A#1} (19#2)\ }
\def\nc#1,#2,{{\sl Nuovo Cim.\/ }{\bf #1} (19#2)\ }
\def\np#1,#2,{{\sl Nucl.\ Phys.\/ }{\bf B#1} (19#2)\ }
\def\pl#1,#2,{{\sl Phys.\ Lett.\/ }{\bf #1B} (19#2)\ }
\def\pla#1,#2,{{\sl Phys.\ Lett.\/ }{\bf #1A} (19#2)\ }
\def\pr#1,#2,{{\sl Phys.\ Rev.\/ }{\bf #1} (19#2)\ }
\def\prd#1,#2,{{\sl Phys.\ Rev.\/ }{\bf D#1} (19#2)\ }
\def\prl#1,#2,{{\sl Phys.\ Rev.\ Lett.\/ }{\bf #1} (19#2)\ }
\def\prp#1,#2,{{\sl Phys.\ Rept.\/ }{\bf #1C} (19#2)\ }
\def\ptp#1,#2,{{\sl Prog.\ Theor.\ Phys.\/ }{\bf #1} (19#2)\ }
\def\ptpsup#1,#2,{{\sl Prog.\ Theor.\ Phys.\/ Suppl.\/ }{\bf #1}
(19#2)\ }
\def\rmp#1,#2,{{\sl Rev.\ Mod.\ Phys.\/ }{\bf #1} (19#2)\ }
\def\yadfiz#1,#2,#3[#4,#5]{{\sl Yad.\ Fiz.\/ }{\bf #1} (19#2) #3%
\ [{\sl Sov.\ J.\ Nucl.\ Phys.\/ }{\bf #4} (19#2) #5]}
\def\zh#1,#2,#3[#4,#5]{{\sl Zh..\ Exp.\ Theor.\ Fiz.\/ }{\bf #1}
(19#2) #3%
\ [{\sl Sov.\ Phys.\ JETP\/ }{\bf #4} (19#2) #5]}

\begin{document}

\begin{titlepage}

\begin{flushright}
CERN-TH/99-308\\
hep-th/9910090
\end{flushright}
\vfil\vfil

\begin{center}

{\Large {\bf Small Instantons and Weak Scale String Theory\\
}}

\vfil

Karim Benakli and Yaron Oz

\vfil

Theory Division, CERN\\
CH-1211, Geneva 23, Switzerland

\end{center}

\vspace{5mm}

\begin{abstract}
\noindent 
We consider heterotic string compactifications to four dimensions
when instantons  shrink to zero size.
If the standard model gauge group originates from the new
gauge symmetry associated with the small instantons singularity,
then  the weakly or strongly coupled 
heterotic string scales can be taken to be arbitrarily low. 
The $SO(32)$ and $E_8\times E_8$ gauge groups can  then 
be very weakly coupled even at the string scale and behave  
as non-abelian global symmetries. 
We comment on a possible role of small instantons
in supersymmetry breaking.

\end{abstract}

\vfil\vfil\vfil
\begin{flushleft}
October 1999
\end{flushleft}
\end{titlepage}

\newpage
\renewcommand{\baselinestretch}{1.05}  

\section{Introduction}

The string scale, compactification scale, and Planck scale
and the relations between
them, are of 
great importance 
for understanding the dynamics of  string theory
and its phenomenological applications.
These relations depend strongly on the choice of the string vacuum \cite{Witten}.
In this note we consider heterotic string compactifications to four dimensions
when instantons  shrink to zero size \cite{SO32,SW,GH} in weakly and strongly coupled \cite{HW} limits of heterotic strings. We discuss some aspects of 
the resulting four-dimensional $\cN=1$ supersymmetric gauge theories.

One of the main aims of our study is to illustrate how the 
heterotic string scale can be made arbitrarily low. Lowering the compactification scale \cite{antoniadis}, the
string scale \cite{lykken} and quantum gravity scale \cite{ADD} to  TeV or 
to intermediate energies \cite{KB,BIQ} allows new perspectives on phenomenological applications of string theory. It was argued that this is possible 
for Type I string theory  in \cite{AADD}, for
the Ho\v rava--Witten  compactifications of M-theory in \cite{KB} and 
for Type II string theory  in \cite{AP} (see also \cite{RS}).

The paper is organized as follows.
In the next section we will briefly discuss heterotic string compactification.
In section 3 we will consider  the singularities when $SO(32)$ instantons shrink to zero size 
and discuss phenomenological aspects of the four-dimensional
$\cN=1$ supersymmetric gauge theories. 
We show that if the standard model gauge group originates from the new
gauge symmetry associated with the small instantons singularity,
then both the weakly and strongly coupled 
heterotic string scales can be taken to be arbitrarily low. 
The $SO(32)$ gauge group can   
be very weakly coupled even at the string scale and play the role of  
a non-abelian  global symmetry.
In section 4 we will consider  the singularities when $E_8\times E_8$ 
instantons shrink to zero size. 
We comment on a possible role of the small instantons
in supersymmetry breaking.

\section{Heterotic string compactification}

Consider $E_8\times E_8$ and the $SO(32)$ heterotic string theory with a 
ten--dimensional string coupling  $\lambda_H = \exp(\phi_H)$ 
and a string scale $l_H$.
At low energies
the effective ten--dimensional description is in terms
of  a super Yang-Mills theory coupled to supergravity.
The ten dimensional gauge coupling $g_{10}$ and the gravitational
coupling $\kappa_{10}$ are 
\beq 
g_{10}^2 = \frac{1}{4}\lambda_H^2l_H^6,~~~~~~\kappa_{10}^2= 
\frac{1}{8}\lambda_H^2l_H^8 \ .
\eeq  
We will be interested in a compactification of
weakly  coupled heterotic string theory on a Calabi--Yau
3-fold. Concretely, we will consider Calabi--Yau 3-folds 
 of the form of a $K3$ fibration
over a $P^1$ base. 
The four-dimensional gravitational 
coupling $\kappa_{4}$ is related to the Newton constant 
$G_N$ \footnote{$G_N \equiv 1/M_p^2 \equiv l_p^2$ where $M_p \sim 10^{19}$ GeV is the 
four-dimensional Planck scale.}
by $\kappa_{4}^2 = 8 \pi G_N$. The latter is given by
\beq
G_N  = \frac {\lambda_H^2 l_H^8}{64 \pi 
\left<V_{K3} V_{P^1}\right>} \ ,
\label{newtonH}
\eeq
where $ V_{K3}$ and  $V_{P^1}$ are the 
volumes of the $K3$ and of the $P^1$ base spaces
respectively. The brackets stand for the average over the compact 
space. 
The tree-level four-dimensional gauge coupling constant 
of an unbroken subgroup of $SO(32)$ and $E_8 \times E_8$ is given by:

\beq
\alpha_H \equiv \frac{g_4^2}{4\pi} = \frac{\lambda_H^2 l_H^6}
{16 \pi f \left<V_{K3} V_{P^1}\right>} \  ,
\label{coupH}
\eeq
with $f$ standing for the 
different normalization of the traces in the 
gauge kinetic term. Below we ignore the model dependence arising from
the factor $f$
and take $f=1$.
These lead to:
\beq
\alpha_H=  4 \frac{l_p^2}{l_H^2}\ .
\label{badH}
\eeq
As is clear from  (\ref{badH}),
requiring a gauge coupling $\alpha_H$ of order 1
implies that $l_H$ is of order $l_p$. In \cite{bachas} 
it was suggested to use one-loop--modified relations with the inclusion of 
threshold correction due to $d$ large dimensions with size $R$. In such a case, 
while $\lambda_H$ governs the strength of gravitational interactions, the 
gauge interaction are governed by:
\beq
\alpha_H^{\rm one-loop}  = \frac {\alpha_H}{1+c\alpha_H (R/l_H)^d}
=  4 \frac{l_p^2}{l_H^2 (1+4 c \frac{l_p^2 R^d}{l_H^{2+d}})} \ ,
\label{loop}
\eeq
with $c$ a constant containing the beta-function coefficient. We see that 
for large $l_H$ a coupling of order 1 is obtained if the denominator is
correspondingly small. Here $R$ is bounded to be smaller than $\sim {\rm TeV}^{-1}$ 
and 
$d =1, 2$ for the supersymmetric case and up to 6 for the non-supersymmetric
 case. This leads roughly to bounds of a string scale higher than  
$\sim 10^{11}$ GeV in the first case and $\sim 10^7$ GeV in the second one.

In the following we 
use the fact that (\ref{coupH}) does not apply to gauge sectors of
{\it weakly} coupled heterotic string that do not descend 
from the ten-dimensional $SO(32)$ or $E_8 \times E_8$ gauge groups.
Such gauge sectors arise from singularities in the moduli space of the
compactification where
extra massless particles are present and there is an extra gauge symmetry.
We will
consider the enhanced gauge symmetries associated 
with the singularities when $SO(32)$\, \footnote{In six dimensions this has been considered in \cite{lykken}.} or $E_8 \times E_8$
instantons shrink to zero size.  
This phenomenon cannot be seen in the conformal field theory description,
but can be seen 
at a weakly coupled heterotic string description.

\section{$SO(32)$ Small Instantons}

Consider first the case of the 
weakly  coupled  $SO(32)$ heterotic string theory compactified
on a $K3$ fibration
over a $P^1$ base. 
We will work in the adiabatic limit of a large $P^1$ base,
which simplifies the discussion
but is not essential.
In this limit we can use a local description as a compactification
to six dimensions on the $K3$ fibre. 
In order to specify a  compactification of the  heterotic string
on a manifold $M$, we have to choose a gauge bundle $V$. One requirement from the compactification
is that
\beq 
\frac{1}{2} p_1(V) = \frac{1}{2} p_1(M) \ ,
\label{cond}
\eeq
where $p_1$ is the first Pontryagin number.
For $K3$ it means that we have to choose the gauge bundle to have
instanton number $24$.
Witten argued \cite{SO32} that at the singularity, associated with a collapse
of $k$ instantons at the same point in $K3$, a new $Sp(k)$ gauge symmetry appears.
In addition   
massless hypermultiplets appear. They consist of $({\bf 32},{\bf 2k})$
of the $SO(32) \times Sp(k)$ gauge group and a massless hypermultiplet in the antisymmetric
representation of $Sp(k)$, which is a singlet of $SO(32)$.

The gauge coupling of the new gauge group is different from the $SO(32)$ gauge coupling.
In order to determine the new gauge coupling it is useful to consider the 
dual picture of the instantons collapse
in Type I string theory at strong coupling.
The analogue in the Type I picture of the collapse of $k$ instantons at the same point 
is $k$ coinciding five-branes. The enhanced gauge group is the five-branes
world-volume gauge group and the massless hypermultiplets are the matter content
of the world-volume gauge theory.
They are obtained by analysing the zero modes of the open strings attached
to the D5-branes \cite{SO32}.
The six-dimensional $Sp(k)$ gauge coupling is $g_{D5}^2 = (2\pi)^3\lambda_Il_I^2$,
where $\lambda_I = \exp(\phi_I)$ is the ten-dimensional Type I string coupling
and $l_I$ is the Type I string scale.
The duality map between Type I string theory and the heterotic string theory
is $\lambda_I = 1/\lambda_H,~~~l_I^2 = \lambda_H l_H^2$.
Thus, the six-dimensional gauge coupling of the small instanton
sector at the heterotic side is given by
$g_{SI}^2=(2\pi)^3 l_H^2$.

We would now like  to reduce the six dimensional $\cN=1$ supersymmetric gauge theory on the $P^1$
and obtain a four-dimensional $\cN=1$ gauge theory.
On the  Type I side this means wrapping the D5 branes on the base $P^1$.
The spectrum of the four-dimensional theory can be obtained as in \cite{Kachru}.
Here we have to to take into account
the non-trivial fibration of the fibre $K3$ over the base $P^1$, which leads to a twist of
the Dirac operator on $P^1$. The reduction preserves one covariantly constant
spinor, yielding $\cN=1$ supersymmetry in four dimensions.
The $Sp(k)$ vector fields survive the reduction as well as some of the
six-dimensional matter hypermultiplets. For instance when $k=1$, $Sp(1) \sim SU(2)$
the 32 hypermultiplets in the fundamental representation
of the $SU(2)$ gauge group give rise to four doublets in four dimensions  \cite{Kachru}.

The 
four-dimensional gauge coupling reads 
\beq
\alpha_{Sp} \ =\ \frac{2 \pi^2 l_H^2}{V_{P^1}} \ ,
\label{sp}
\eeq
where $V_{P^1}$ is the volume of the base.
 The configuration where one identifies the standard model gauge group with 
the small instantons gauge sector will allow us to consider arbitrary low  heterotic 
$SO(32)$ string scale.

There are three dimensionless expansion parameters in the system that we require
to be small in order for the weakly coupled description to be valid.
The first is the expansion parameter of the perturbative string description
in ten dimensions ${\lambda_H^2}/{\left( 2 \pi \right)^5}$ \cite{Kap}: 
\beq
\frac {\lambda_H^2}{\left( 2 \pi \right)^5} =\frac {2}{\pi^4 } 
\frac { l_p^2}{ l_H^2}\frac {\left< V_{K3} V_{P^1}\right>}{\lambda_H^6} \ ,
\label{st}
\eeq
which we require to be smaller
than 1 in order for the heterotic string to be weakly coupled
in space-time.
The second parameter is  $\frac {l_H^6}{\left< V_{K3} V_{P^1}\right>}$, which
we require to be smaller
than 1 in order for the heterotic string to be weakly coupled
on the world-sheet.
The third parameter is $\alpha_{Sp}$ in (\ref{sp}), which
we require to be smaller
than 1 in order for the new gauge symmetry to be weakly coupled.

Let us now analyse these conditions for the validity of the weakly coupled
description.
We choose $K3$ such that  $\frac {l_H^4}{\left< V_{K3}\right>} < 1$ 
so that the small
instanton picture is valid. 
We assume that
$\left< V_{K3} V_{P^1}\right> \sim \left< V_{K3}\right> \left<  V_{P^1}\right> $.
Together with the requirement that $\alpha_{Sp} \sim \frac{l_H^2}{V_{P^1}}$ 
(\ref{sp}) be small, it guarantees that $\frac {l_H^6}{\left< V_{K3} V_{P^1}\right>}$
is small too.
Finally, in order for $\frac {\lambda_H^2}{\left( 2 \pi \right)^5}$
to be small, we require that $\frac { l_p^2}{ l_H^2}$ be small, namely a 
weakness of
gravitational interactions is consistent with the weakly coupled description.

We can view the weakness of
gravitational interactions as arising either from a large $K3$ volume
or from a very small string coupling constant.  For instance, taking $\alpha_{Sp} \sim 1/10$ 
as a rough estimate, the first possibility
 arises, with a choice:\\ 
$ \lambda_H \sim 1$ \, and \, $\left< V_{K3}\right>^{1/4}\sim 10,\ \,  10^{3},\ \,  
10^{6} \,  l_H$ \, 
for \, $l_H^{-1} \sim 10^{16},\ \,  10^{11},\ \, 10^{4}$ GeV respectively.
The second possibility arises, with a choice:\\
$\left< V_{K3}\right>^{1/4} \sim  {\rm few}~~l_H$ \, and \, $ \lambda_H \sim 10^{-1},\,  10^{-6},\, 10^{-13}$,  for  $l_H^{-1} \sim \, 10^{16}, \ \,  10^{11}, \ \,  10^{4}$ GeV 
respectively.

The above discussion is valid till the gauge coupling (\ref{sp}) 
is large $\alpha_{Sp} \sim \frac {l_H^2}{\left< V_{P^1}\right>} \sim 1$ 
and we cannot trust a perturbative analysis. 
Passing to the dual Type I description is not useful either, since
at these energies the  
heterotic $SO(32)$ gauge coupling  
$\alpha_{SO}=  4 \frac{l_p^2}{l_H^2}$ is still weak 
and therefore the $SO(32)$ gauge coupling on the Type I side is large.
At this energy scale the system in no longer four-dimensional and 
we probe the physics of six dimensions.

If $\lambda_H$ is chosen to be very small, at energies below the string scale,
 the unbroken part of the
 $SO(32)$ symmetry is very weakly coupled and it is seen from the $Sp(k)$ side as a non-abelian ``global'' symmetry. Such kinds of symmetries can
 be useful for phenomenological issues such as forbidding
operators that could lead to proton decay or other exotic processes. On the other hand the gravitational interactions are still weak at the string scale. The main experimental signature would be the observation of effects due to the Kaluza--Klein modes of $P^1$. If one instead explains the weakness
of gravitational interactions by a large $K3$ volume (as in Type I scenarios)
then at energies of order $l_H^{-1}$ the $SO(32)$ symmetry coupling is of the 
same order as the one of $Sp(k)$ and cannot be viewed as a global symmetry. This is due to the sum of the contributions from the Kaluza--Klein states propagating in the $K3$.
Moreover at the string scale the gravitational interactions are now of the same strength as  the gauge ones.

A large class of models with various gauge groups and matter content for which the above
discussion continues to hold can be obtained by shrinking instantons at ADE singularities
of $K3$ \cite{aspin}. On the Type I side these models are obtained by placing
the five-branes at these singular points \cite{BI,Intri}.
In these models the gauge groups are products of the classical gauge groups
$\prod_{i,j,k} SO(n_i) \times Sp(m_j) \times U(l_k)$ arranged according to 
quiver (moose)  diagrams related to the extended Dynkin diagrams of the ADE groups.

Finally we note that instead of a compactification on the base $P^1$, we can reduce 
to four dimensions on two circles with non-trivial boundary conditions 
on  the circles.
This leads to $\cN=0$ gauge theories in four dimensions. The above discussion continues
to hold at energies $E \ll 1/R$, where $R$ is the radius of the circles.
At higher energies we will probe these two compact coordinates, by producing
the associated Kaluza--Klein states.

\section{$E_8\times E_8$ Small Instantons}

Consider now the case of $E_8\times E_8$ heterotic string compactified 
on a $K3$ fibration
over a $P^1$ base, 
in the adiabatic limit.
Denote by $n_1,n_2$ the instanton numbers of the two
$E_8$ groups.
We have to choose the gauge bundle with 
$n_1+n_2=24$.
When we shrink some of the instantons to zero size we do not get a new gauge symmetry
in six dimensions. Instead, we get massless tensor multiplets and hypermultiplets
in six dimensions \cite{SW,GH}. 
The six-dimensional tensor multiplet contains a 2-form field $B_{\mu\nu}$
which is self-dual $dB =*dB$. 
In the dual picture of M-theory compactified on
$S^1/Z_2$, this process is viewed as  placing M5-branes near one of the $E_8$ walls.
There are tensionless strings that arise from membranes stretched between the
M5-branes and the $E_8$ wall and couple to $B$.
When we reduce on $P^1$ the tensor multiplets do  not give rise to gauge fields
but rather to matter multiplets. This is due to the fact that there are no 1-forms
$\omega$ on $P^1$, which otherwise would enable us to decompose
$dB = F \wedge \omega$ and obtain the gauge field strength $F$.

We can however obtain vectors fields in six dimensions
and a large class of gauge groups and matter content by shrinking $E_8$ instantons 
at ADE singularities \cite{aspin}.
For instance, if we shrink $k$ instantons at $A_{n-1}$ singularity
we get a gauge group $\prod_{i=2}^{n-1}SU(i)\times SU(n)^{k-2n+1}\times
 \prod_{j=2}^{n-1}SU(j)$
with bi-fundamental matter.
The six-dimensional gauge couplings of these gauge groups is determined by vacuum
expectation values (vev's) $\left < \phi \right >$ 
of scalars in particular tensor multiplets \cite{aspin}.
These
 scalars in six dimensions have dimension two 
and we can choose  vev's $\left < \phi \right > \sim 1/l_H^2$.
Upon reduction on $P^1$ we can identically repeat 
the discussion in the previous section for the weakly coupled heterotic 
strings case. For the Ho\v rava--Witten  compactifications an arbitrarily low scale can be obtained by taking  all or some of the five dimensions transverse 
to the M5-brane large.

\section{Discussion}

We have argued that the tree-level gauge and gravitational
couplings  dependence on the string and compactification scales allow
the latter to be  arbitrarily low. In addition to the necessity  for
building realistic models, many important questions related to the
dynamics of string theory remain  to be addressed.
For instance: How  is  supersymmetry
broken?, Does this allow the size of the couplings  to be  small or the
volumes large, as required for lower values of the string scales?  and
How do loop corrections modify our analysis?

Here we would like to briefly comment on some issues of supersymmetry 
breaking. An interesting possibility is to use the small instantons to break
supersymmetry. For this we choose a gauge bundle with $(n_1,-n_2)$
instanton number such that $n_1-n_2=24$ and shrink the $n_2$
anti-instantons.  This configuration breaks suspersymmetry
completely. On the strongly coupled  heterotic side,
 described by M-theory compactified on
$S^1/Z_2$, this would
correspond to placing $n_2$ anti-M5-branes on one of the
walls. Supersymmetry is then broken at the eleven-dimensional Planck
scale. This would need to be at TeV if we live on the
anti-M5-branes, at intermediate energies if we live on the opposite
wall, and  somewhere in between if we live on the same wall
(gauge mediation). This is the heterotic theory 
scenario  corresponding to the proposal of \cite{antibrane} for type I
vacua. Obviously the issue of stability has to be addressed in these
models.

Another possibility  is to break 
supersymmetry spontaneously by wrapping the M5-brane on the boundary wall around
a  non-supersymmetric cycle \cite{Yaron} (the use of five-branes in the bulk
as hidden sector to break supersymmetry was suggested in \cite{ovrut}). This 
is useful, for 
instance, in scenarios with a non-standard embedding \cite{nse} 
where the volume on the hidden wall of the Calabi--Yau space  is large, making 
the gauge interaction on the wall weaker than needed 
 to induce non-perturbative effects 
that are able  to break supersymmetry at 
desired scales. Also when  small instantons are localized at different points 
of $K3$, they may act as hidden sectors as in the F-theory scenario of \cite{KL}.  
The observable and ``hidden'' sectors communicate through both gravity and the $SO(32)$ or $E_8 \times E_8$ gauge symmetries.
Finally, for very low values of the string scale, one may replace in
 all models discussed above 
the base $P^1$ of the Calabi--Yau 3-fold by two circles with 
boundary conditions that break all the supersymmetries.

To summarize, 
we considered generic features of the gauge theories arising 
from small instantons and pointed out to some aspects that are of
phenomenological relevance, such as a 
possibility to lower the string scale, 
extremely weakly coupled gauge symmetries that act as global non-abelian symmetries,
and finally a possible role  in the dynamics of supersymmetry breaking. 
These, we believe, deserve further studies.

\section*{Acknowledgement}
We would like to thank P. Mayr  for useful discussions.

\newpage



\begin{thebibliography}{99}



\bibitem{Witten} E. Witten,``Strong Coupling Expansion of Calabi-Yau
Compactification'',
{\em Nucl.Phys.} {\bf B471} (1996) 135,  hep-th/9602070.


\bibitem{SO32} E. Witten, ``Small Instantons in String Theory'',
{\em Nucl.Phys.} {\bf B460} (1996) 541, hep-th/9511030.



\bibitem{SW} N. Seiberg and  E. Witten, ``Comments on String Dynamics in
Six Dimensions'',
{\em Nucl.Phys.} {\bf B471} (1996) 121, hep-th/9603003.


\bibitem{GH}  O. Ganor and  A. Hanany,
``Small $E_8$ Instantons and Tensionless Non-Critical Strings'',
{\em  Nucl.Phys.} {\bf  B474} (1996) 122,
hep-th/9602120.



\bibitem{HW}
P. Ho\v rava and E. Witten, ``Heterotic and Type I String Dynamics from
Eleven Dimensions'',  
 {\em Nucl.Phys.} {\bf B460} (1996) 506, hep-th/9510209;
``Eleven-Dimensional Supergravity on a Manifold with Boundary'', {\em
Nucl.Phys.} {\bf B475} (1996) 94, hep-th/9603142.

\bibitem{antoniadis} I. Antoniadis, ``A Possible New Dimension at a Few TeV'',
    {\em  Phys.Lett.} {\bf  B246} (1990) 377.


\bibitem{lykken}  J.D.~Lykken, 
``Weak Scale Superstrings'', {\em  Phys.Rev.} {\bf  D54} (1996) 3693,
 hep-th/9603133.


\bibitem{ADD} N.~Arkani-Hamed, S.~Dimopoulos and G.~Dvali, 
``The Hierarchy Problem and New Dimensions at a Millimeter'', {\em
Phys.Lett.} {\bf B429}
(1998) 263,  hep-ph/9803315.


\bibitem{KB} K. Benakli, ``Phenomenology of Low Quantum Gravity Scale
Models'',
{\em Phys.Rev.} {\bf  D60}
(1999) 104002, hep-ph/9809582.

\bibitem{BIQ} C.P. Burgess, L.E. Iba\~nez and F. Quevedo,
``Strings at the Intermediate Scale, or is the Fermi Scale Dual to the
Planck Scale?'', {\em Phys.Lett.} {\bf B447} (1999) 257, hep-ph/9810535.

\bibitem{AADD} 
I.~Antoniadis, N.~Arkani-Hamed, S.~Dimopoulos and G.~Dvali,
``New Dimensions at a Millimeter to a Fermi and Superstrings at a TeV'', 
{\em Phys.Lett.} {\bf B436} (1998) 257,
hep-ph/9804398; for model building see for example: G.~Shiu and S.-H.H.~Tye,
``TeV Scale Superstring and Extra Dimensions'', {\em Phys.Rev.} {\bf  D58}
(1998) 106007, hep-th/9806143;
G. Aldazabal, L.E. Iba\~nez and  F. Quevedo, ``Standard-Like Models with
Broken Supersymmetry from Type I String Vacua'', hep-th/9909172.



\bibitem{AP} I. Antoniadis and B. Pioline, ``Low-Scale Closed Strings and
their Duals'',  {\em Nucl.Phys.} {\bf B550} (1999) 41, hep-th/9902055.

\bibitem{RS} L. Randall and R. Sundrum, ``A Large Mass Hierarchy from a
Small Extra Dimension'', hep-ph/9905221.

\bibitem{bachas} C. Bachas, private communication.

\bibitem{Kachru} S. Kachru, N. Seiberg and E. Silverstein,
``SUSY Gauge Dynamics and Singularities of 4d N=1 String Vacua'',
{\em Nucl.Phys.} {\bf B480} (1996) 170.

\bibitem{Kap} E. Caceres, V. S. Kaplunovsky and I. M. Mandelberg,
``Large Volume String Compactifications, Revisited'', {\em Nucl.Phys.}
{\bf B493} 73, hep-th/9606036.

\bibitem{aspin}  P. S. Aspinwall and D. R. Morrison, ``Point-like Instantons on K3 Orbifolds'', 
{\em Nucl.Phys.} {\bf B503} (1997) 533,  hep-th/9705104.


\bibitem{BI} J. D. Blum and  K. Intriligator,
`` New Phases of String Theory and 6d RG Fixed Points via Branes at
 Orbifold Singularities'', {\em Nucl.Phys.} {\bf B506} (1997) 199,  
hep-th/9705044.


\bibitem{Intri} K. Intriligator, ``New String Theories in Six Dimensions 
via Branes at Orbifold Singularities'', hep-th/9708117; ``Compactified 
Little String Theories and Compact Moduli Spaces of Vacua'', hep-th/9909219.

\bibitem{antibrane} I. Antoniadis, E. Dudas and  A. Sagnotti, ``Brane Supersymmetry Breaking'',  hep-th/9908023; G. Aldazabal and A. M. Uranga, ``Tachyon-Free Non-Supersymmetric Type IIB Orientifolds via Brane-Antibrane Systems'', hep-th/9908072; G. Aldazabal et {\em al.},  last refrence in [11]. 

\bibitem{Yaron} J. de Boer, K. Hori, H. Ooguri and Y. Oz, ``Branes and 
Dynamical 
Supersymmetry Breaking'', {\em Nucl.Phys.} {\bf B522} (1998) 20, 
hep-th/9801060; A. Sen, ``BPS D-branes on Non-Supersymmetric Cycles'', {\em JHEP} {\bf 9812} (1998) 021, hep-th/9812031. 

\bibitem{ovrut} A. Lukas, B. A. Ovrut and D. Waldram, ``Five-branes and 
Supersymmetry Breaking in M-Theory'', {\em JHEP} {\bf 9904} (1999) 009, 
hep-th/9901017.

\bibitem{nse} K. Benakli, ``Scales and Cosmological Applications of M-Theory'',
{\em Phys.Lett.} {\bf B447} (1999) 51, hep-th/9805181; 
S. Stieberger, ``(0,2) Heterotic Gauge Couplings and  Their
 M-Theory Origin'', {\em  Nucl.Phys.} {\bf B541} (1999) 109, hep-th/9807124.
 

\bibitem{KL} V. Kaplunovsky and J. Louis, `` Phenomenological Aspects of F-Theory'', {\em Phys.Lett.} {\bf  B417} (1998) 45.
 
\end{thebibliography}
\end{document}